\documentclass{aa}
\usepackage{natbib}
\bibpunct{(}{)}{;}{a}{}{,} % to follow the A&A style
\usepackage{graphicx}
\usepackage{aalongtable}
\usepackage{txfonts}
\begin{document}
  \title{The spiral structure of our Milky Way Galaxy}
  \author{L.~G. Hou
    \inst{1,2}
    \and J.~L. Han
    \inst{1}
    \and W.~B. Shi
      \inst{1,3}
  }
  
   \offprints{J.L. Han: hjl@bao.ac.cn}
   
   \institute{National Astronomical Observatories, Chinese Academy of Sciences,
     Jia-20, DaTun Road, Chaoyang District, Beijing 100012, China
             \and
             Department of Physics, School of Physics, Peking
             University, Beijing 100871, China
             \and
             Department of Space Science and Applied Physics,
             Shandong University at Weihai, 180 Cultural West Road,
             Shandong 264209, China
             }

   \date{}

% \abstract{}{}{}{}{}
% 5 {} token are mandatory

  \abstract
{The spiral structure of our Milky Way Galaxy is not yet known. HII regions
and giant molecular clouds are the most prominent spiral tracers. Models
with 2, 3 or 4 arms have been proposed to outline the structure of our
Galaxy. }
{Recently, new data of spiral tracers covering a larger region of the
Galactic disk have been published. We wish to outline the spiral structure
of the Milky way using all tracer data.}
{We collected the spiral tracer data of our Milky Way from the literature,
namely, HII regions and giant molecular clouds (GMCs). With weighting
factors based on the excitation parameters of HII regions or the masses of
GMCs, we fitted the distribution of these tracers with models of two,
three, four spiral-arms or polynomial spiral arms. The distances of
tracers, if not available from stellar or direct measurements, were
estimated kinetically from the standard rotation curve of Brand \& Blitz
(1993) with $R_0$=8.5 kpc, and $\Theta_0$=220 km s$^{-1}$ or the newly
fitted rotation curves with $R_0$=8.0 kpc and $\Theta_0$=220 km s$^{-1}$ 
or $R_0$=8.4 kpc and $\Theta_0$=254 km s$^{-1}$.
}
{We found that the two-arm logarithmic model cannot fit the data in many
regions. The three- and the four-arm logarithmic models are able to connect
most tracers. However, at least two observed tangential directions cannot be
matched by the three- or four-arm model. We composed a polynomial spiral arm
model, which can not only fit the tracer distribution but also match
observed tangential directions. Using new rotation curves with $R_0$=8.0 kpc
and $\Theta_0$=220 km s$^{-1}$ and $R_0$=8.4 kpc and $\Theta_0$=254 km
s$^{-1}$ for the estimation of kinematic distances, we found that the
distribution of HII regions and GMCs can fit the models well, although the
results do not change significantly compared to the parameters with the
standard $R_0$ and $\Theta_0$.}
{}


   \maketitle
%
%________________________________________________________________

\section{Introduction}

\label{sect:intro}

The Milky Way galaxy is known to be a spiral galaxy, but its
detailed spiral structure has not been well revealed. The Milky Way
may have two arms, or three arms, or even more complicated structures
\citep{val08}. At present, the number of arms and arm parameters
(i.e. the pitch angle and the { initial Galactocentric radius}) have
not been well determined.  Nevertheless, some consensus on the
Galactic structure has been reached \citep[][hereafter R03]{rus03}:
(1) the tangential directions of the spiral arms have been identified
from the maxima of the thermal radio continuum and molecular emission
(see Table 1 in \citeauthor{eg99}, \citeyear{eg99} or Table 2 in
\citeauthor{val08}, \citeyear{val08}); (2) the Sagittarius arm and the
Carina arm are linked as a single arm; (3) the Sun lies between the
Perseus arm and Sagittarius arm.

To outline the structure of our Galaxy, one should use all reliable tracers
with accurate distances. Primary tracers are:

(1) HII regions. HII regions are the birthplace of young stars. They
are clouds of atomic hydrogen ionized by bright young stars. Their
radio emission cannot  be attenuated by dust extinction, and
therefore they can be detected even in distant parts of the Galactic
plane. \citet{gg76} investigated the distribution of the available
sample of HII regions that time, and outlined four segments of
arms. \citet{dwbw80} observed more HII regions in the first quadrant
and \citet{ch87} reported comprehensive data of southern HII regions
in the Galaxy over the Galactic longitude $l=210\degr$ to
$l=360\degr$, which help to delineate the spiral structure, especially
for the Carina arm and the Crux arm.

(2) Giant molecular clouds (GMCs). The GMCs are vast assemblies of molecular
gas with a mass of $10^4\sim10^6 M_{\odot}$ and a size of tens of pc. They
have an average density of $10^2\sim10^3$~cm$^{-3}$ but their core density
can reach $10^6\sim10^7$~cm$^{-3}$. Because the interstellar medium is
almost optically thin for the CO emission line, distant clouds in the
Galactic plane can be detected. \citet{cdt86} showed that the GMCs are good
tracers of the Carina arm. \citet{dect86} solved the distance ambiguity of
some GMCs in the first quadrant, and used them to outline the Sagittarius
arm. \citet{srby87} studied warm molecular clouds in the first quadrant, and
delineated three arm-like structures, the Perseus arm, the Sagittarius arm
and another possible one, the Scutum arm. \citet{gcbt88} cataloged GMCs over
the Galactic longitude from $l=270\degr$ to $l=300\degr$, and outlined the
Sagittarius-Carina arm using GMCs over 23~kpc in the Galactic plane.  More
GMCs should be identified from the complete CO map of our Galaxy composed by
\citet{dht01}, but unfortunately, this has not yet been done.

In addition, HI gas is one of the major components of the interstellar
medium in spiral galaxies. \citet{lbh06} decomposed the HI data into the
smoothed component and perturbed surface density map, and found a spiral
structure out to at least 25~kpc. Four arm-segments appear in the outer
Milky Way disk.

As shown in external Galaxies, the distribution of the brightest HII regions
and of the most massive molecular clouds generally traces the grand design
structure. To our knowledge, previous authors examining the spiral structure
of the Milky Way used only HII regions or GMCs, and none considered all
tracers together for spiral structure, except for R03 who considered the
complexes of HII regions and molecular clouds. R03 showed the distribution
of a large sample of star-forming complexes (constituted by HII regions,
ionized patches and/or molecular clouds or the mixture of any two), and
fitted it with two, three, or four logarithmic spiral arm models. The
four-arm model is slightly more preferable than the three-arm model. We
collect a catalog of HII regions and GMCs, and use those tracers together
to show the grand design of our Galaxy.  In Sect.~2, we discuss spiral
tracers and determine their parameters (distance, and the assigned weight -
here we mean the weighting factor rather than the mass). In section 3, we
present the distribution of the tracers and discuss the models. Conclusions
are presented in Sect.4.

%__________________________________________________________________

\section{Tracer data for the Galactic spiral structure}
\label{sect:Obs}

To reveal the structure of our Galaxy, tracers spread over the whole
Galactic disk are needed. Also, their distances have to be determined.

\subsection{Tracers for the Galactic spiral structure: data}

For any tracer in the {\it outer} Galaxy, if its stellar distance cannot be
determined, one can use the kinematic method to estimate the
distance. However, for a tracer in the inner Galaxy, the distance ambiguity
is a problem. Two possible distances correspond to the same observed radial
velocity. Because the distance ambiguity of HII regions can be resolved by
HI/$H_2CO$ emission/absorption observations or the HI self-absorption method
\citep[e.g.][]{ab09,tl08}, HII regions are excellent tracers in the inner
Galaxy.

In addition to \citet{gg76}, \citet{dwbw80} and \citet{ch87} on the spiral
structure of the Milky Way using HII regions as spiral tracers,
\citet{pbd+03} collected a catalog of Galactic HII regions, which contained
1442 sources. \citet{pdd04} presented a new, detailed analysis of the
spatial distribution of some Galactic HII regions. \citet{ahck02},
\citet{was+03}, and \citet{swa+04} reported simultaneous $H110\alpha$ and
$H_2CO$ line observations toward HII regions in the first quadrant and
resolved the distance ambiguity. \citet{kb94}, \citet{kjb+03},
\citet{frwc03}, \citet{pmg08}, \citet{ab09} used HI absorption to resolve
the distance ambiguity for other samples of HII regions. R03 established a
star-forming complex catalog, and derived Galactic structure. \citet{rag07}
revised distances of 32 HII regions.

For GMCs, besides \citet{dect86}, \citet{srby87} and \citet{gcbt88} for
spiral structure of the Milky Way, there are also investigations on Galactic
molecular clouds. \citet{mk88} reported 31 clouds in the first quadrant
outside the solar circle. \citet{dbt90} observed 32 clouds, related to the
Outer arm in the first quadrant.  \citet{sod91} studied 35 clouds located in
the second, third and fourth quadrants. \citet{mab97} identified 177 clouds
in the third quadrant, and shown disk structure as well as warping of the
disk.  \citet{css90} studied 18 molecular clouds in the outer Galactic
disk. \citet{bw94} detected a sample of molecular clouds located in
the far outer Milky Way. \citet{nomf05} revealed 70 molecular clouds in the
Galactic Warp with a kinematic distance greater than about 14.5
kpc. \citet{hcs01} summarized the properties of molecular regions in the
outer galaxy from the Five College Radio Astronomy Observatory Outer Galaxy
Survey.

We have collected the published data of HII regions and GMCs from the
references above, including position, velocity, flux, etc. The stellar
distances of HII regions are used when possible, otherwise the kinematic
distances are estimated by using three rotation curves (see below), one with
the constants $R_0=8.5$ kpc and $\Theta_0=$220 km~s$^{-1}$, another with
$R_0=8.0$ kpc and $\Theta_0=$220 km~s$^{-1}$ and also the new one with
$R_0=8.4$ kpc and $\Theta_0=$254 km~s$^{-1}$. Here $R_0$ is the distance
between the Sun and the Galactic Center, and $\Theta_0$ is the velocity of
the Sun circling around the Galactic Center. The excitation parameters of HII
regions and the masses of GMCs are estimated { and rescaled to the adopted
distances.}  Clouds with a mass less than $10^4 M_{\odot}$ are not massive
and will not be considered. We have cross-identified the tracers according
to their longitudes, latitudes and velocities to avoid redundancy. We list
all related parameters of the tracers with references in Table~\ref{tab_a1}
and Table~\ref{tab_a2} (online version only).

\subsection{Rotation curves and distances of tracers}

The distance of a given spiral tracer could be best determined by
triangulation observations 
\citep[e.g.][]{xrzm06,xrm+08,brm+08,zzr+08,rmb+08,mrm+08}. Only a few
clouds or HII regions have been so well measured. For many tracers,
the distance of the associated bright stars is adopted from the
literature. Here, we will use these measurements to verify the
rotation curves.

The Galactic rotation curve has been derived from the tracers with
their stellar distances and velocities already determined. In the last 
20 years, four rotation curves have
been used for the spiral structure of the Milky Way: one for the whole
Galaxy from \citet[][hereafter BB93]{bb93}, the two for the north part
of the Galaxy from \citet{cle85} and \citet{fbs89}, and the very
simple flat rotation curve with $\Theta=$ 220 km~s$^{-1}$ for the
whole Galaxy. R03 obtained a Galactic rotation curve by using his
sample of star-forming complexes, which is almost the same as that of
BB93.

Using the sources with stellar distances in our collected sample, we
verified the four possible rotation curves mentioned above with the IAU
standard values, $R_0=8.5$ kpc and $\Theta_0=220$ km~s$^{-1}$. Similarly to
R03, we calculated the parameter $\tauup$:
\begin{equation}
    \tauup=\frac{1}{N} \sum_{i=1}^{N}
    \sqrt{
    \frac{(R_i-R_t)^2}{\sigma_{R_i}^2}+
    \frac{(\omegaup_i-\omegaup_t)^2}{\sigma_{\omegaup_i}^2}
         }
  .
\label{eq1}
\end{equation}
Here, $R_{i}$ and $\omegaup_{i}$ are the Galactocentric distance and the
angular rotation velocity of a tracer, respectively, and $\sigma_{R_{i}}$
and $\sigma_{\omegaup_{i}}$ are their uncertainties. $R_{t}$, $\omegaup_{t}$
are the coordinates of the theoretical points. The most appropriate rotation
curve should minimize the parameter $\tauup$.  We found that the rotation
curve of \citet{cle85} can indeed minimize the parameter $\tauup$ for the
north part and the whole Galaxy with a polynomial fit. However it fits all
the structures of the curve, even those caused by local motions. The
$\tauup$ for the rotation curve of \citet{fbs89} and the simple flat one are
larger than that of BB93 both in the north part and whole Galaxy. The
rotation curve of BB93 (i.e. Eq.~\ref{curve} below with a=1.00767, b=0.0394
and c=0.00712) is reasonable for all data of the whole Galaxy. We will
directly adopt it to estimate the kinematic distances of the spiral tracers
for the case with $R_0$=8.5~kpc and $\Theta_0$=220 km~s$^{-1}$.
\begin{figure}
\centering\includegraphics[width=0.47\textwidth]{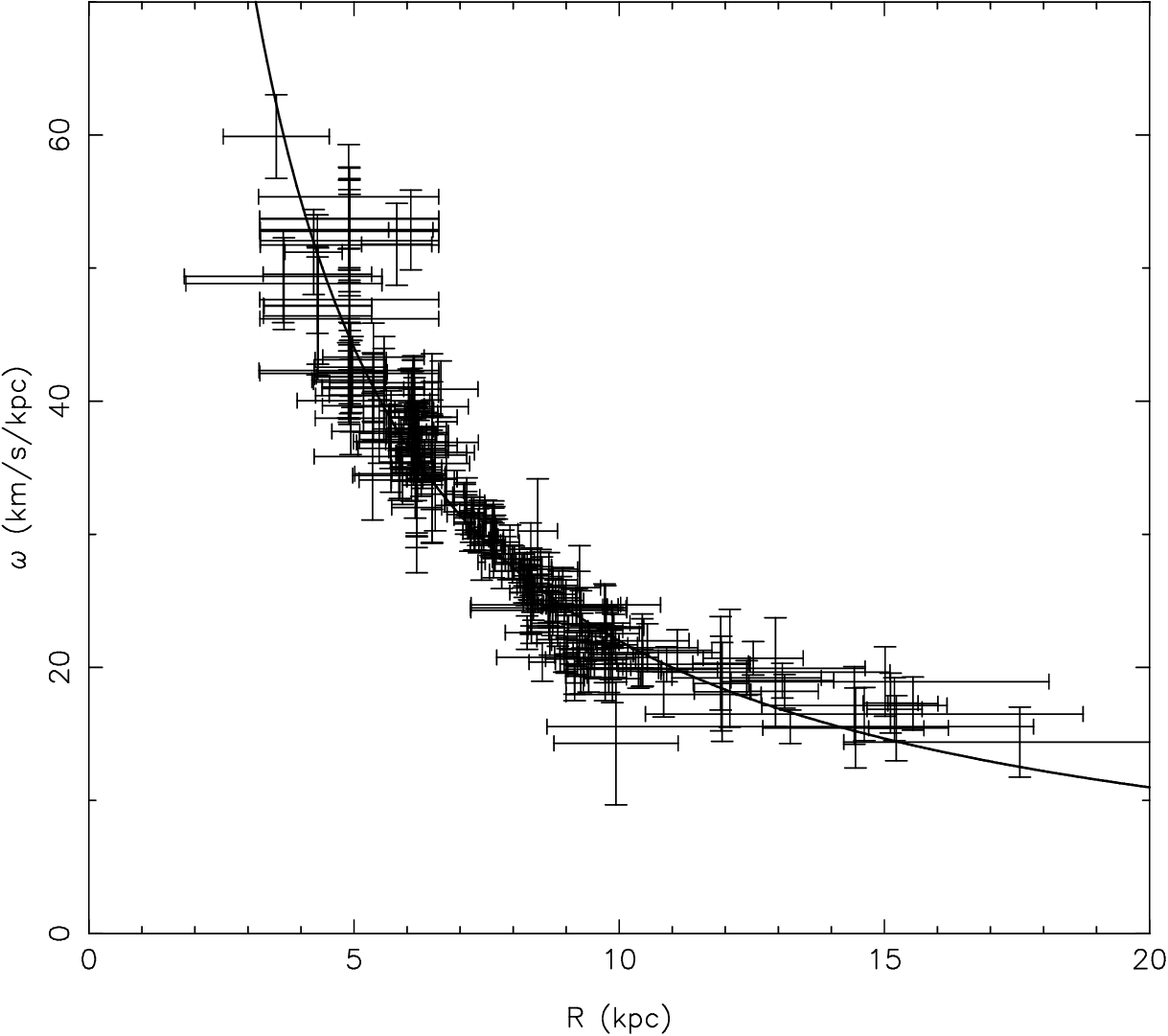}
\caption{The rotation curve fitted with solar parameters $R_0=8.0$ kpc and
$\Theta_0=220$ km~s$^{-1}$. Data are taken from our collected sample listed
in Table~\ref{tab_a1} and Table~\ref{tab_a2}.}
\label{rotation}
\end{figure}
\begin{figure}
\centering \includegraphics[angle=270,width=0.47\textwidth]{9692f2a.ps}
\centering \includegraphics[angle=270,width=0.47\textwidth]{9692f2b.ps}
\caption{The distribution of excitation parameters of the HII regions
({\it top panel}) and masses of molecular clouds ({\it bottom
panel}). The rotation curve with $R_0$=8.5 kpc and $\Theta_0$=220
km~s$^{-1}$ was used if the distances of these tracers were estimated
kinetically.}
\label{fig1}
\end{figure}
%%%%%%%%
\begin{figure*}
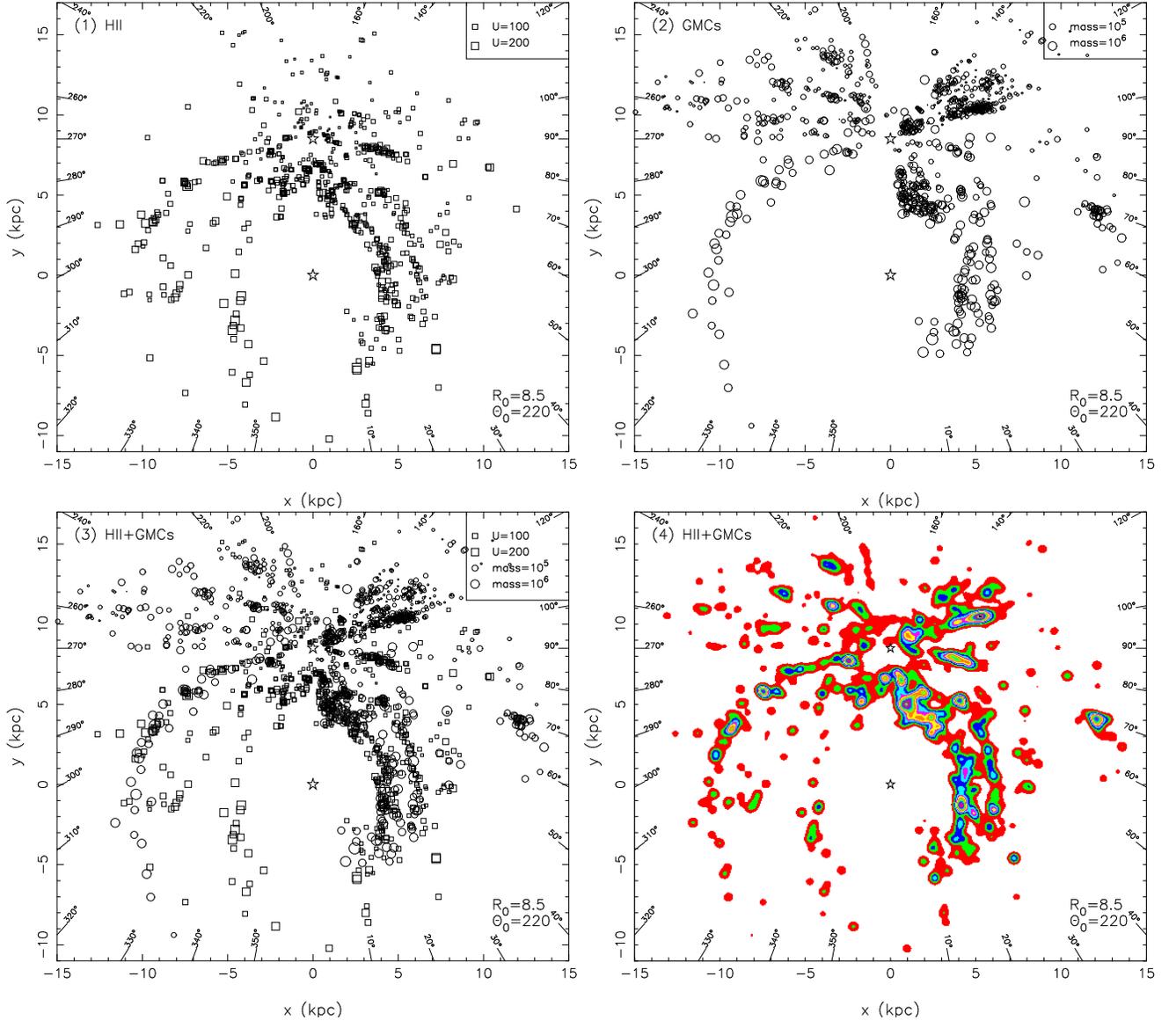

\centering\includegraphics[width=0.47\textwidth]{9692f3a.ps}
\centering\includegraphics[width=0.47\textwidth]{9692f3b.ps}
\centering\includegraphics[width=0.47\textwidth]{9692f3c.ps}
\centering\includegraphics[width=0.47\textwidth]{9692f3d.ps}
\caption{ {\it Panel (1)} is the distribution of HII regions, {\it
    Panel (2)} is the distribution of GMCs, and {\it Panel (3)} is the
  distribution of HII regions and GMCs together for illustration of
  the Galactic spiral structure. The solar parameters $R_0$=8.5 kpc
  and $\Theta_0$=220 km s$^{-1}$ were adopted.  The coordinates
  originate from the Galactic center, and the Sun is located at
  ($x=0.0$~kpc, $y=8.5$~kpc). The open squares indicate the HII
  regions with the symbol area proportional to exciting
  parameters. The open circles indicate GMCs with the symbol size
  proportional to $log(M_{GMCs})$ (see Sect.~2.3). {\it Panel (4)} is
  the color distribution of both kinds of tracers, each brightened as
  a Gaussian with the amplitude of the weighting parameter, $B$,
  so that the spiral arms are clearly demonstrated.}
\label{distri}
\end{figure*}

More and more pieces of evidence of $R_0$=8.0 kpc and $\Theta_0$=220 km
s$^{-1}$ have been found \citep{reid93,gsw+08, esg+03, gub08,get+09}. The
change of these constants may affect the kinematic distances of tracers and
then affect the derived structure of the Milky Way. Therefore, we fit a new
rotation curve with \citep[see ][R03]{bb93}:
\begin{equation}
\omega/\omega_0=a(R/R_0)^{b-1}+c(R_0/R)
\label{curve}
\end{equation}
We set the rotation curve to pass through the new $R_0$ and $\Theta_0$, and
set $c=1-a$.  By minimizing $\tau$ in Eq.~\ref{eq1}, we found the best
values, $a=1.59073$, $b=-0.000816$. The result is shown in
Fig.~\ref{rotation}. In Section 3.3, we will use this rotation curve,
and also the newly determined one with $R_0$=8.4~kpc and $\Theta_0$=254
km~s$^{-1}$ by \citet{rmz+09}, to estimate the kinematic distances and
related parameters, and outline the spiral structure. Note that
\citet{get+09} obtained $R_0 = 8.33\pm0.35$~kpc from the combined fitting to
observations of 16 years for stellar orbits around the black hole in the
Galactic center.

\subsection{Tracers of Galactic spiral structure: weights}

The brighter the HII region, the better it is as a tracer of spiral
structure. We use the excitation parameters of HII regions as a weighting
factor to demonstrate spiral structure. \citet{gg76} delineated four
arm-segments using HII regions with an excitation parameter $U$ greater than
70 pc~cm$^{-2}$. R03 used the excitation parameter as a weighting factor in
their fitting process. The excitation parameter $U$ (in pc~cm$^{-2}$) is
defined as \citep{sm69}:
\begin{equation}
 U=4.5526\alpha(\nuup,T)^{-1}\nuup^{0.1}T^{0.35}S_{\nuup}D^{2}
\label{eq:LebsequeII}
\end{equation}
Here, $T$ is the temperature (in K), $S_{\nuup}$ is the radio flux density
(in Jy), $\nuup$ is the frequency (in GHz), $D$ is the distance of the
tracers to the Sun (in kpc), and $\alpha(\nuup,T)$ is a parameter close to 1
\citep[][]{sm69}. We collected the $S_{\nuup}$ from the literature when
available. The excitation parameter $U$ is then estimated via Eq.(3) using
$S_{\nuup}$ and assuming $T=8500$~K. For a temperature in the range of 7000
$\sim $10000~K the value of excitation parameter differs by less than about
$6\%$. The distribution of excitation parameters of HII regions in our
sample is shown in Fig.~\ref{fig1}. It ranges from 5 to 332 pc~cm$^{-2}$
with a peak around 50 pc~cm$^{-2}$.

Massive molecular clouds are more concentrated to in the spiral arms than
clouds with low masses. The mass of a molecular cloud $M_{GMCs}$ was
rescaled with newly adopted distances. The distribution of $M_{GMCs}$ is
shown in Fig.~\ref{fig1}, ranging from 1$\times 10^4 M_{\odot}$ to
2.5$\times 10^7 M_{\odot}$. More clouds have a lower mass.

In order to {\it use the HII regions and GMCs together} to outline the
spiral structure of our Galaxy, a reasonable match of weighting parameters
should be considered to measure their relative importance to the spiral
structure. Obviously, a HII region with a higher excitation parameter or a
GMC with a larger $M_{GMCs}$ should have a larger value of weight. For the
HII regions, we take the weighting parameter $B$=$U$/(100 pc~cm$^{-2}$), so
that an HII region with $U$=100 pc cm$^{-2}$ has a weight of 1 in the
determination of the spiral pattern, while an HII region with $U$=200 pc
cm$^{-2}$ has a weight of 2. For the GMCs, the weighting parameter is
taken to be $B=\log(M_{GMCs}/10^4M_{\odot})$, so that the clouds with a mass
of $10^5 M_{\odot}$ have a weight of $B$=1, and clouds of $10^6 M_{\odot}$
have a weight of 2. We have also tried other weighting measures, such as
$B$=$U$/(50 pc~cm$^{-2}$) and $B$=2log($M_{GMCs}$/$10^4M_{\odot}$), and
found that the arm structure is similar but with more emphasis on GMCs.

An additional factor needed in the spiral fitting is the distance
uncertainty of each tracer. If the uncertainty of observed stellar distance
is known, we use it directly. If the kinematic distance is used, we { adopt
a systemic velocity uncertainty of $\pm $5km s$^{-1}$}, and then the
distance uncertainty can be derived by using a rotation curve. As we search
for the best match of spiral arms in X-Y coordinates, distance uncertainty
will be expressed in X and Y. We set a fitting weighting factor
$w_x=0.5/\sigma_x$ and $w_y=0.5/\sigma_y$. We assign all tracers with a
distance accuracy better than 0.5 kpc (i.e. $\sigma_x< 0.5$ kpc ) to have
$w_x=1$, and $w_y=1$ if $\sigma_y < 0.5$ kpc.

\section{The spiral arms of the Milky Way}

Fig.~\ref{distri} shows the distribution of tracers for the Galactic
spiral structure projected onto the Galactic plane. The sizes of
symbols have been scaled according to the weighting factors $B$
obtained from excitation parameter $U$ or cloud mass.
GMCs in a large region of the fourth Galactic quadrant have been
mapped by \citet{dht01}, but unfortunately, the discrete
GMCs have not been identified and explicitly listed in the
literature.
To show spiral arms better using available data of HII regions and
GMCs, in the fourth panel of Fig.~\ref{distri}, we use a Gaussian
function to { brighten each tracer} through:
\begin{equation}\label{eq:LebsequeI}
        L(x,y)=\sum_i {\frac{B(i)} {\sqrt{2\pi
    \sigma^2}}} \exp(-{\frac{(x_{i}-x)^{2}+(y_{i}-y)^{2}} { 2
    \sigma^2}})
\end{equation}
Here we took $\sigma$=0.2, which means that each tracer will brighten an
area with a radius of about 200 pc. This is larger than the size of a GMC,
but we can delineate the spiral structure more clearly. A different value of
$\sigma$ in a suitable range (e.g. 50~pc -- 400~pc) produces a similar
figure. In the fourth Galactic quadrant, three arm-like structures exist in
this region. The Carina arm ($l\sim290\degr$) is very clear. The other two
arm-like structures correspond to the Crux ($l\sim318\degr$) and the Norma
arm ($l\sim333\degr$), respectively. In the first quadrant, both HII regions
and GMCs also seem to show three arm-like structures, the Sagittarius arm in
the middle ($l\sim40\degr$), the Scutum arm inside ($l\sim25\degr$) as well
as the Perseus arm on the outer side ($x\sim8$~kpc, $l\sim60\degr$). These
arms are clear in \citet{was+03} and \citet{swa+04} from HII region
distributions and in \citet{srby87} from the distribution of molecular
clouds. There may be another arm-like segment in the far outer Galaxy traced
by GMCs. In the second and third Galactic quadrants, no obvious structure
can be identified from the distribution of the main tracers.

\begin{figure}
\centering\includegraphics[width=0.46\textwidth]{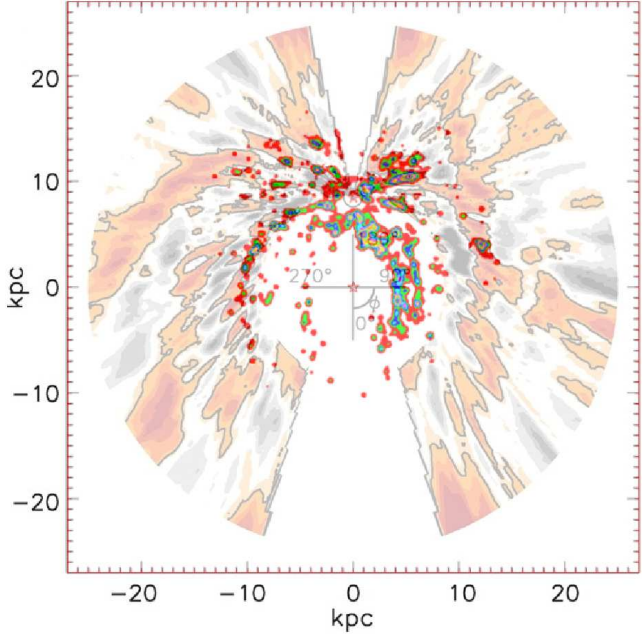}
\caption{The spiral structure of HII regions and GMCs ($R_0=8.5$ kpc
  and $\Theta_0=220$ km s$^{-1}$) is overlaid on the HI map
  \citep{lbh06}.}
\label{onHI}
\end{figure}

\begin{figure}
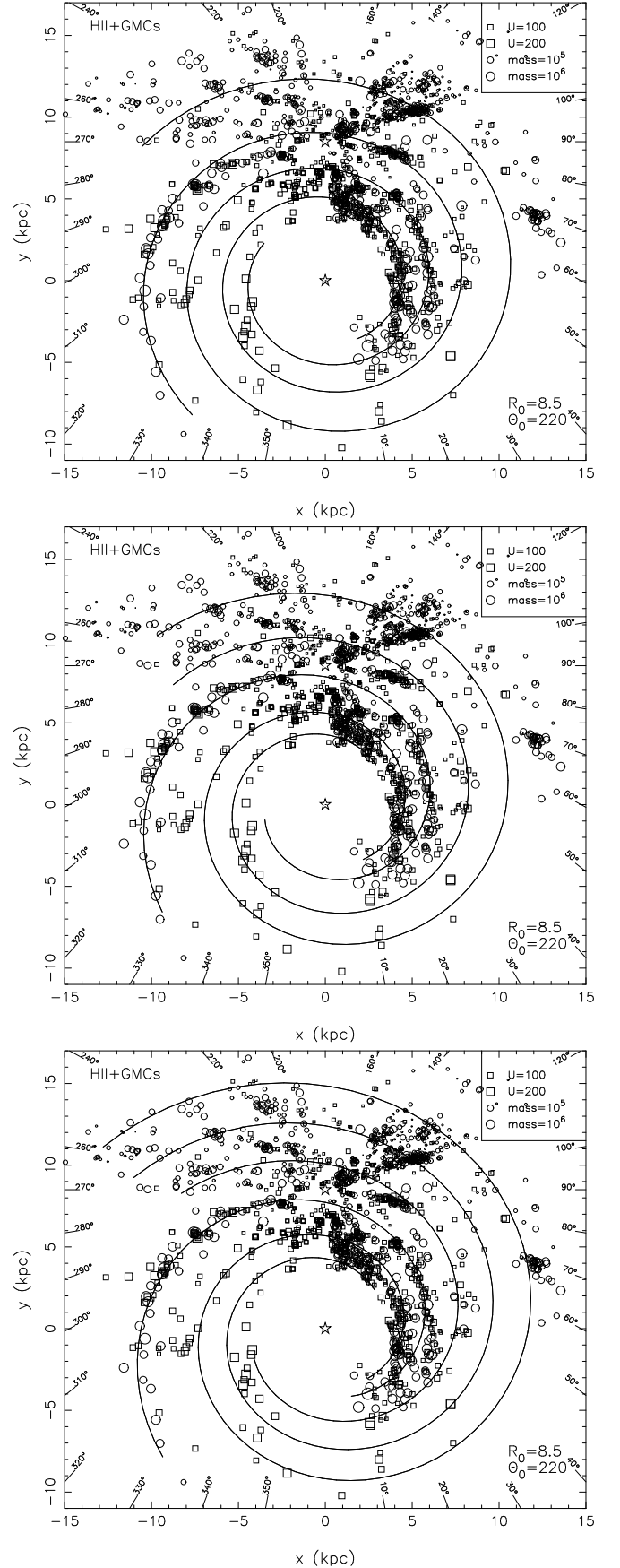

\begin{tabular}{cc}
\mbox{\includegraphics[width=0.47\textwidth]{9692f5a.ps}} \\
\mbox{\includegraphics[width=0.47\textwidth]{9692f5b.ps}} \\
\mbox{\includegraphics[width=0.47\textwidth]{9692f5c.ps}} \\
\begin{minipage}[t]{0.48\textwidth}
%\vspace{-0.5\textwidth}
\caption{The distribution of HII regions and GMCs ($R_0=8.5$ kpc and
$\Theta_0=220$ km s$^{-1}$) and the best-fitting two-arm model ({\it 
top panel}), three-arm model ({\it middle panel}) and four-arm 
model ({\it bottom panel}).}
\label{234model}
\end{minipage}
\end{tabular}
%\vspace{-10mm}
\end{figure}

In Fig.~\ref{onHI}, we overlay our data onto the HI map of \citet{lbh06},
which was obtained { with the solar parameters of $R_{0}=8.5$~kpc and
$\Theta_0=220$~km s$^{-1}$}. In the HI distribution, there are 3 obvious
arm-like structures in the third and fourth quadrants. The Carina arm
overlaps with the most inner arm-like structure in the HI distribution, and
the Perseus arm may be linked to the intermediate arm-like structure, which
can be matched in our 3-arm, 4-arm, and polynomial logarithmic arm
models. The counterpart for the outer arm-like structure in the HI map is
difficult to identify, but most likely there are one or two spiral arms
beyond the Perseus arm.

It is not clear whether the spiral structure of the Milky Way is of
logarithmic nature. Canonically, the two, three or four logarithmic arms
have been used to fit the tracer data and to model the grand design of our
Galaxy. We will try to fit these conventional models to the data, and we
also present the polynomial logarithmic arm model with varying pitch angle.
In addition to fitting the model to the tracer distribution, it is important
to compare the tangential directions of modelled spiral arms to the observed
tangents \citep[see][]{eg99,bro92,val08}.

In Section 3.3, we also show the distribution of tracers for spiral arms by
taking the kinematic distances estimated by using the rotation curves with
two sets of solar parameters, $R_{0}=8.0$~kpc and $\Theta_{0}=220$~km
s$^{-1}$ and the recent determination of $R_{0}=8.4$~kpc and
$\Theta_{0}=254$~km s$^{-1}$ from \citet{rmz+09}.

\subsection{The logarithmic arm models}

The logarithmic-arm model describes the $i$-th spiral arm with:
\begin{equation}
\ln\frac{r}{R_{i}}=(\theta-\theta_{i})\tan\psiup_{i} ;
\end{equation}
Here, $r$, $\theta$ are the polar coordinates centered at the Galactic
center, $R_{i}$ is the initial radius (in kpc) at the start azimuthal
angle $\theta_{i}$ for the $i$-th arm, and $\psiup_{i}$ is the pitch
angle of the arm. The $\theta$ starts at the positive x-axis and
increases counterclockwise. We consider the expression: 
\begin{equation}\label{eq:LebsequeI}
        Z=\frac{1}{\sum{B_i}}\sum
    B_{i}\sqrt{{(x_{i}-x_{t})^{2}} {w_{x_i}^2} +
    {(y_{i}-y_{t})^{2}} {w_{y_i}^2}}
\end{equation}
for all tracers; here $x_{i}$ and $y_{i}$ are Cartesian
coordinates of the tracer, $x_{t}$ and $y_{t}$ are the
coordinates of the nearest point from all spiral arms to the tracer.
$w_{x_i}$ and $w_{y_i}$ are weighting factors for the location
uncertainty of $x_{i}$ and $y_{i}$, respectively, and $B_{i}$ is the
weighting factor related to the excitation parameters of HII regions
or the masses of GMCs. The best spiral arm model should be able to
minimize the parameter $Z$.

We assume that each arm always has its own pitch angle in the 2-, 3-, 4-arm
models and fit them to the data \citep[see also][]{ns07}. By searching for
the best parameters of each arm, $R_{i}$, $\theta_{i}$ and $\psiup_{i}$ in
reasonable ranges, to find the minimized $Z$, we obtained the best spiral
arm models as listed in Table~\ref{234para}. Three best-fitting models
together with the data are plotted in Fig.~\ref{234model}. The tangential
directions of arms in all models are listed in Table.~\ref{tan}

\begin{table*}
\caption{The parameters of the best fitting models: the initial radius,
$R_{i}$, the start azimuthal angle, $\theta_{i}$, and the pitch angle, 
$\psiup_{i}$, for the $i$-th arm.}
\begin{center}
\begin{tabular}{clclclclclclcc}
\hline
\hline
model & $R_{1}$&$\theta_{1}$& $\psiup_{1}$ & $R_{2}$ & $\theta_{2}$ &
$\psiup_{2}$ & $R_{3}$ & $\theta_{3}$ & $\psiup_{3}$ & $R_{4}$ &
$\theta_{4}$& $\psiup_{4}$& $Z$  \\
  & (kpc)&$(^{\circ})$& $(^{\circ})$ & (kpc) & $(^{\circ})$ &
$(^{\circ})$ & (kpc) & $(^{\circ})$ & $(^{\circ})$  &
$(^{\circ})$& $(^{\circ})$& (kpc)& \\
\hline
\multicolumn{10}{l}{For distribution of tracers with $R_{0}=8.5$~kpc and $\Theta_0=220$~km s$^{-1}$} \\
\hline
2-arm model & 4.2 & 147& 5.3 & 4.0 & 297 & 5.2& & & & & & &0.47 \\
3-arm model & 3.8 & 40 & 7.8 & 3.6 & 195 & 9.9 & 4.0 & 303 & 7.5& & & &0.45  \\
4-arm model & 3.7 & 40 & 9.6 & 4.5 & 205 & 10.7 & 4.4 & 290 & 11.4 & 3.9 &
309 & 8.7 & 0.35 \\
\hline
\multicolumn{10}{l}{For distribution of tracers with $R_{0}=8.0$~kpc and $\Theta_0=220$~km s$^{-1}$} \\
\hline
2-arm model & 4.0 & 151& 5.1 & 3.9 & 315 & 5.0& & & & & & &0.43 \\
3-arm model & 3.6 & 40 & 8.0 & 3.5 & 195 & 9.6 & 3.9 & 303 & 7.4& & & &0.41  \\
4-arm model & 3.6 & 40 & 9.2 & 3.7 & 205 & 12.5 & 4.2 & 290 & 11.1 & 3.8 &
309 & 8.4 & 0.33 \\

\hline\hline
\end{tabular}
\end{center}
\label{234para}
\end{table*}

\begin{table*}
\caption{Tangential directions from observations \citep[][]{bro92,eg99} and
  the best fitting models.}
\begin{center}
\begin{tabular}{ccccccc}
\hline
\hline
model & Scutum & Sagittarius & Carina & Crux & Norma & 3-kpc \\
 & $(^{\circ})$ & $(^{\circ})$ & $(^{\circ})$ & $(^{\circ})$& $(^{\circ})$ & $(^{\circ})$ \\
\hline
observed mean value& 25,31 & 51 & 284 & 310 & 327 & 339  \\[1mm]
\hline
\multicolumn{6}{l}{For distribution of tracers with $R_{0}=8.5$~kpc and $\Theta_{0}=220$~km s$^{-1}$} \\
\hline
2-arm model & 33 & 49 &  &301 & 319 & 330 \\
3-arm model & 26,36& 58 & 284 & 313 & 325 &   \\
4-arm model &25,36 & 55 & 283 & 311 & 323 &  \\
polynomial model&27,33  &53  &284  & 309 & 329 &  \\
\hline
\multicolumn{6}{l}{For distribution of tracers with $R_{0}=8.0$~kpc and $\Theta_{0}=220$~km s$^{-1}$} \\
\hline
2-arm model & 33 & 49 &  &303 & 320 & 330 \\
3-arm model & 26,38& 59 & 282 & 311 & 324 &   \\
4-arm model &26,37 & 55 & 284 & 310 & 322 &  \\
polynomial model&27,34  &53  &284  & 309 & 329 &  \\
\hline
\multicolumn{6}{l}{For distribution of tracers with $R_{0}=8.4$~kpc and $\Theta_{0}=254$~km s$^{-1}$} \\
\hline
polynomial model&27,33  &53  &284  & 309 & 329 &  \\
\hline\hline
\end{tabular}
\end{center}
\label{tan}
\end{table*}

The best fitting two-arm logarithmic model places the Sun in the
Carina arm, and it also connects the Local arm and the Perseus arm
with the Carina arm, which is inconsistent with observations. In
fact, the connection of the Sagittarius arm and the Carina arm as a
single arm has been well established from observations. Such a
connection cannot  be obtained in any two-arm model, as found
by R03.  In addition, the tangential directions calculated from
this model are also very different from observed values. The parameter
$Z$ is slightly larger than that of the 3- and 4- arm models. The best
two-arm logarithmic model can be excluded, conclusively.

In the best fitting 3-arm and 4-arm models, the Sagittarius arm and the
Carina arm are connected and look very similar. The Scutum arm in the 1st
Galactic quadrant outlined by HII regions and molecular clouds is connected
to the Crux arm in the 4th Galactic quadrant. This arm is well fitted by
both the 3-arm and the 4-arm models. However, the obvious difference arises
from the extension of the Crux arm. It connects to the Perseus+1 arm in the
3-arm model, but to the Perseus+2 arm in the the 4-arm model.  It is not
clear from current available data whether there is one arm or two arms in
the few kpc outside the Perseus arm. Conservatively, the 3-arm model
predicts only one Perseus+1 arm and no Perseus+2 arm, which may be more
realistic. However, \citet{lbh06} showed that the HI gas distribution
indicates that both the Perseus+1 arm and Perseus+2 arm are possible. The
molecular clouds at $l\sim70\degr$ at 13 kpc \citep{dbt90} may be related to
the Perseus+2 arm. In the GC direction, the Norma arm is connected to the
Perseus arm in the 3-arm model. In the 4-arm model, the Norma arm is
connected to the Perseus+1 arm, and the Perseus arm starts from another arm
inside the Norma arm, corresponding to the Norma-Cygnus arm in the four-arm
model of R03.

The tangential directions of arms in models are listed in Table~2 for
comparison with observations. The double values of the Scutum tangent
correspond to both the Scutum arm and another interior arm segment, which
can be well fitted in our 3-arm and 4-arm models. However an obvious
deviation is seen from the tangent of the actual Scutum arm ($l\sim31\degr$
vs.  $l\sim36\degr$). The tangent of the Sagittarius arm always deviates
from observed values by $7\degr$ or $4\degr$ in the 3-arm or the 4-arm
model. The tangent of the Norma arm differs by $4\degr$ in the 4-arm models.

\begin{table}
\caption{The parameters of the polynomial logarithmic arm model.}
\begin{center}
\begin{tabular}{cccccc}
\hline
\hline
$i$-th arm & $a_i$ & $b_i$ & $c_i$ & $d_i$ & $\theta_i$ \\
\hline
\multicolumn{5}{l}{For distribution with $R_{0}=8.5$~kpc and $\Theta_{0}=220$~km s$^{-1}$}& $(^{\circ})$\\
\hline
arm-1& 1.376 & -0.07792 & 0.04309 & 0 &41 \\
arm-2&7.330 & -2.302 & 0.2849 & -0.01059 &304   \\
arm-3&10.403 & -3.526 & 0.4620 & -0.01895 &298   \\
arm-4& 1.978 & -0.1181 & 0.02098 & 0 &338 \\
arm-5&2.297 &0.09116 &0.04273 & 0 &35\\
$Z$&0.31   &   &    &     &\\
\hline
\multicolumn{5}{l}{For distribution with $R_{0}=8.0$~kpc and $\Theta_{0}=220$~km s$^{-1}$}& $(^{\circ})$ \\
\hline
arm-1& 1.302 & -0.06629 & 0.04115 & 0 &47 \\
arm-2&7.5616 & -2.448 & 0.3090 & -0.01187 &304  \\
arm-3&9.9843 & -3.417 & 0.4522 & -0.01873 &298    \\
arm-4& 1.5189 & -0.00581 & 0.01286 & 0 &338 \\
arm-5&2.1949 &0.1037 &0.03317 & 0 &35 \\
$Z$&0.26   &   &    &   \\
\hline
\multicolumn{5}{l}{For distribution with $R_{0}=8.4$~kpc and $\Theta_{0}=254$~km s$^{-1}$}& $(^{\circ})$ \\
\hline
arm-1& 1.3747 & -0.0737 & 0.04318 & 0 &41 \\
arm-2&11.2880 & -3.8678 & 0.4900 & -0.01949 &304  \\
arm-3&10.2992 & -3.519 & 0.4662 & -0.01937 &298    \\
arm-4& 1.5645 & -0.005055 & 0.0132 & 0 &338 \\
arm-5&2.2973 &0.004478 &0.0750 & 0 &55 \\
$Z$&0.29   &   &    &   \\
\hline\hline
\end{tabular}
\end{center}
\label{5arm}
\end{table}

\begin{figure}
\includegraphics[angle=270,width=0.47\textwidth]{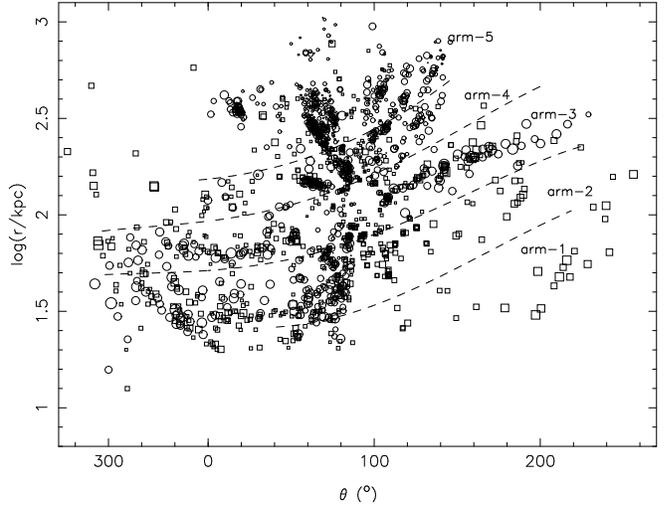} 
\caption{The distribution of tracer data ($R_0=8.5$ kpc and
  $\Theta_0=220$ km s$^{-1}$) plotted in { the $log(r)$-$\theta$
    diagram}. Here $\theta$ starts at the positive $x$-axis and
  increase counterclockwise. The dash lines roughly separate main arm
  tracers.}
\label{r_theta}
\end{figure}

\begin{figure}
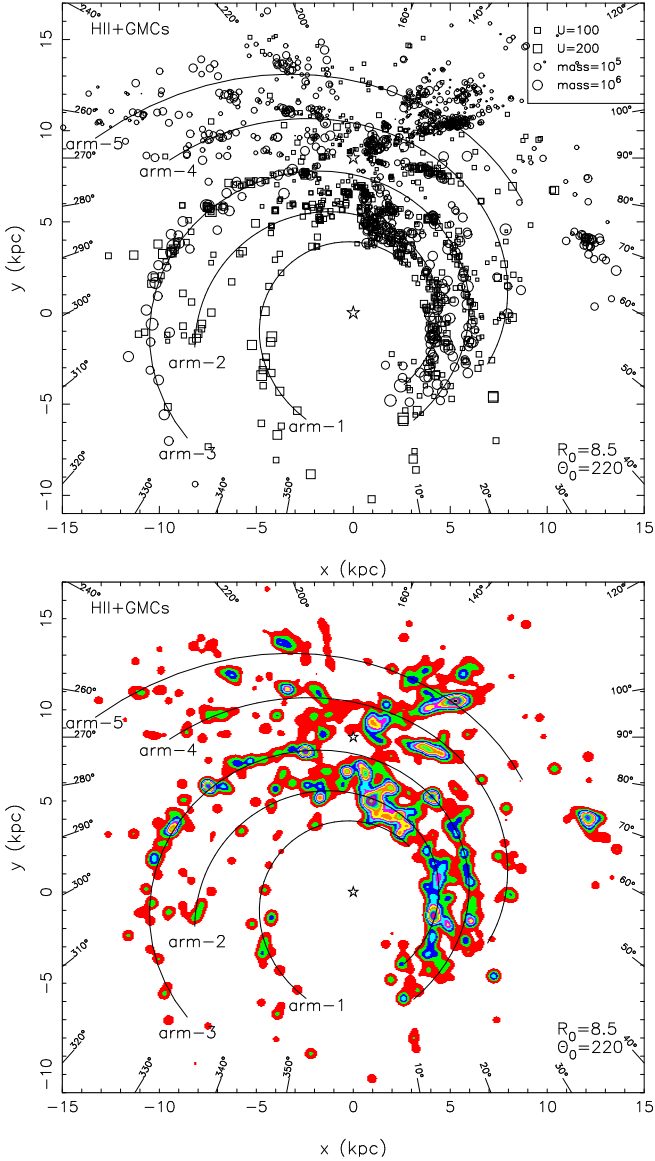

\includegraphics[width=0.47\textwidth]{9692f7a.ps} \\
\includegraphics[width=0.47\textwidth]{9692f7b.ps} 
\caption{The best-fitting polynomial logarithmic-arm model ($R_0=8.5$ kpc
and $\Theta_0=220$ km~s$^{-1}$), plotted onto the data distribution
({\it top panel}) and the color brightened-tracer image ({\it 
bottom panel}) of tracers. Almost all main tracers are connected by
the outlined arms, except for one complex at (x,y)=(12~kpc, 4~kpc)
which either has an overestimated distance or it belongs to another
outer arm.}
\label{p_model}
\end{figure}

\begin{figure}
\includegraphics[angle=270,width=0.47\textwidth]{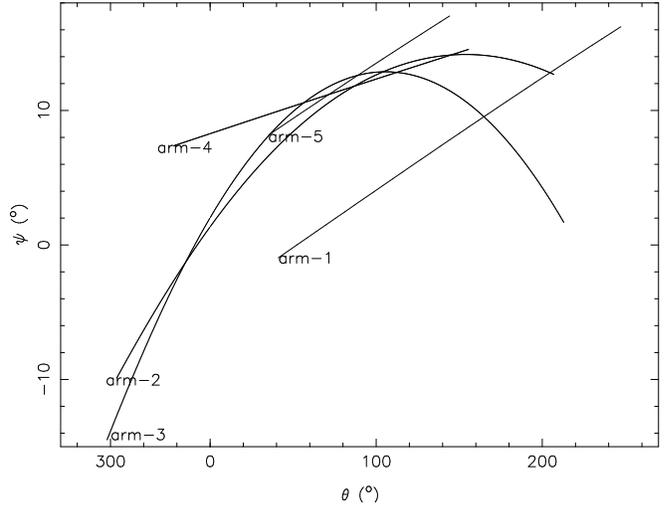}
\caption{The pitch angle ($\psi$) variation of spiral arms in the
accepted polynomial logarithmic-arm model (with $R_0=8.5$ kpc and
$\Theta_0=220$ km s$^{-1}$).}
\label{pitch09} 
\end{figure}

\subsection{The polynomial logarithmic arm model}

Note that these spiral arm models are assumed to be purely
logarithmic. However, the pitch angle of spiral arms can vary along
the arm \citep[see e.g.][]{sj98}. If so, a polynomial logarithmic arm
model may be necessary to fit both the distribution and tangential
directions.

We first plotted the tracer data in the $log(r)$-$\theta$ diagram (see
Fig.~\ref{r_theta}), and roughly identified arm-like features and then
separated the data by the dashed-lines. The Sagittarius-Carina arm is very
prominent in the plot, which obviously separated in the tracer data from the
other groups. The Scutum-Crux arm are also well-isolated in the tracer data,
except for those at small $r$ where the data are mixed with those from the
Sagittarius arm. The innermost arm seems clearly to have a group of tracers,
which in fact is the Norma arm and connected to the Scutum arm. However,
outside the Sagittarius-Carina arm, the data are well mixed, and one can
barely distinguish the Perseus arm and the Perseus+1 arm, which can only be
separated manually in a very rough manner.

The best model should satisfy the observed tangential directions and
also minimize the parameter $Z$. We separate our total sample into
several sub-samples for five arms (see Fig.~\ref{r_theta}), and fit
them separately with the polynomial logarithmic arm model:
\begin{equation}
\ln r = a_i + b_i \theta+c_i\theta^2+d_i\theta^3
\end{equation}
to obtain the initial values of each arm, and finally we fit all arms
together to the whole sample of data. After using the weighting
factors, and minimizing the residual value by Eq.(6), we obtained
the best model with the fitting parameters listed in Table~3. The data
and arms are plotted in Fig.~\ref{p_model}. The tangential directions
of this model are also listed in Table~2.

In this model, almost all tracers are best connected with the outlined
spiral arms. Compared to the pure logarithmic models, the polynomial
logarithmic arm model is more conservative in connecting known tracers. It
does not outline the possible arm trajectory in the regions without tracers
and predict the possible connection between the outer arms and inner
arms. As seen in Fig.~\ref{p_model}, the main spiral structures are well
traced. From the center we find arm-1 to arm-5, which correspond to the
Norma arm, the Scutum-Crux arm, the Sagittarius-Carina arm, the Perseus arm
and the Perseus+1 arm.  Looking at the HI data in Fig.~\ref{onHI}, one may
see that arm-5, i.e. the Perseus+1 arm, seems to be connected to the obvious
outer HI arm outlined by \citet{lbh06} at the both ends at about
(x,y)=($-$12~kpc, +12~kpc) and (+8~kpc, +10~kpc). The Perseus arm is
probably connected to the long HI arm at ($-$12~kpc, +3~kpc). The tangential
directions of this model (Table 2) are more consistent with the
observed values than those from the logarithmic arm model. We conclude that
such a polynomial arm model is a better description of the observed
arm features.

The pitch angles of spiral arms vary in the form of $ tan(\psi_i)=
b_i+2c_i\theta+3d_i\theta^2$, in contrast to the constant pitch angles for
the pure logarithmic arms. The pitch angles vary with azimuthal angle, as
shown in Fig.~\ref{pitch09}. Variable pitch angles have been found in nearby
galaxies. For example, \citet{pd92} used a simple model with a variable
pitch angle to obtain the leading or trailing character of the arms of some
galaxies.  Note, however, that the innermost segments of arm-2
($l<\sim25\degr$) and arm-3 ($l<\sim35\degr$), probably due to the distance
uncertainty of tracers, show unreasonable negative pitch angles in
Fig.\ref{pitch09}, similar to the possible outer HI arm ($x\sim10$~kpc) in
the 1st quadrant as shown in Fig.~\ref{onHI}. This is obtained from
tracer data, which thus needs to be resolved.

\begin{figure}
\includegraphics[width=0.47\textwidth]{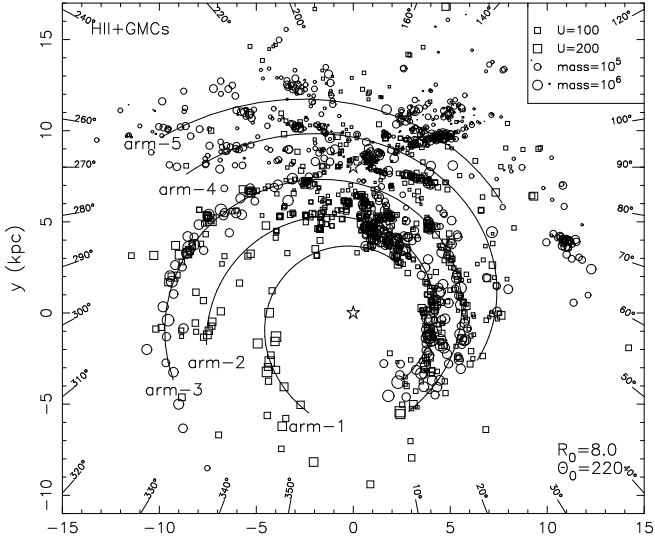}\\
\includegraphics[width=0.47\textwidth]{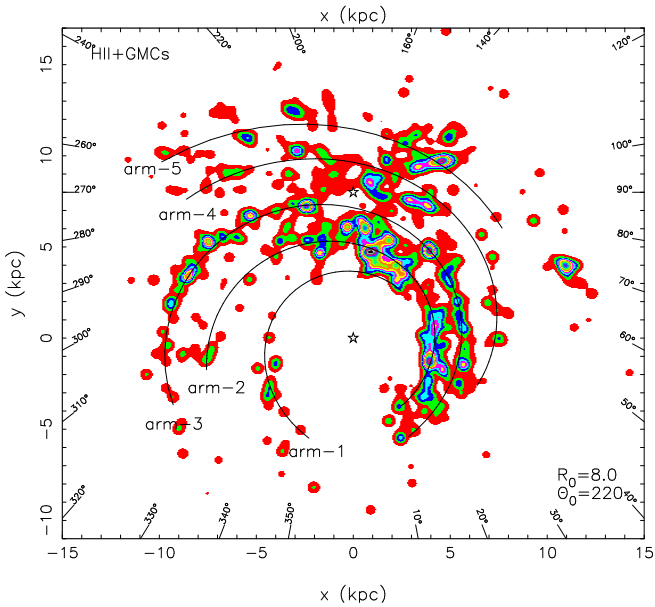}
\caption{The  best-fitting polynomial logarithmic-arm model for
  $R_0=8.0$ kpc and $\Theta_0=220$ km~s$^{-1}$, plotted on the data
  distribution ({\it top panel}) and the color brightened-tracer image
  ({\it bottom panel}) of tracers. }
\label{5arm8.0}
\end{figure}

\begin{figure}
\includegraphics[width=0.47\textwidth]{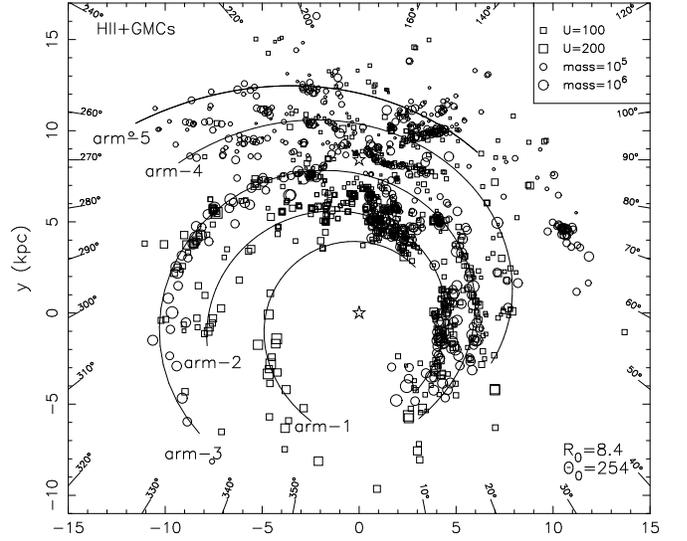}\\
\includegraphics[width=0.47\textwidth]{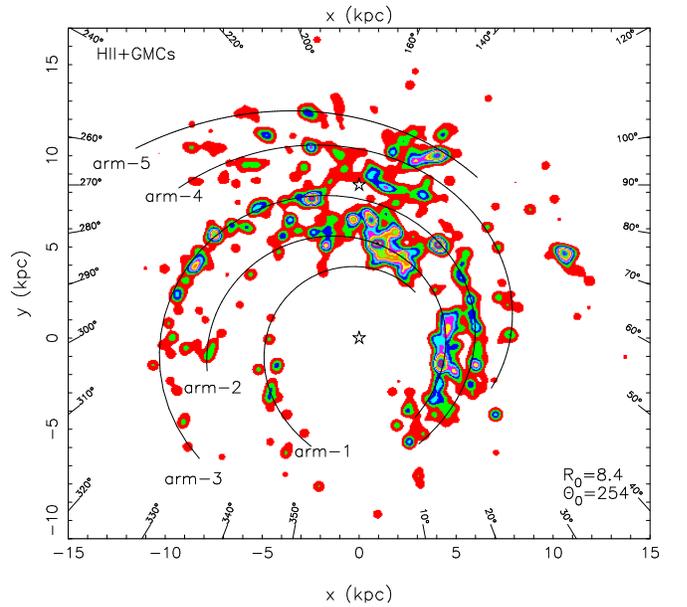}
\caption{The same as Fig.~\ref{5arm8.0} but with the rotation curve
from \citet{rmz+09} with $R_0=8.4$ kpc and $\Theta_0=254$ km~s$^{-1}$.}
\label{5arm254}
\end{figure}

\subsection{The influence of the solar parameters $R_0$ and $\Theta_0$}

We also have re-estimated the distances of HII regions and GMCs, if not
available from other measurements, by using the rotation curve for $R_0=8.0$
kpc and $\Theta_0=220$ km s$^{-1}$. Then the excitation parameters of HII
regions and the masses of GMCs are accordingly recalculated, as listed in
Table~\ref{tab_a1} and Table~\ref{tab_a2}. The final distribution of the
tracers is shown in Fig.~\ref{5arm8.0}. The overall design is similar to
that of $R_0=8.5$ kpc and $\Theta_0=220$ km s$^{-1}$. We also fit the
distribution with two, three, and four spiral-arm models, and the parameter
values from the best models are listed in Table~\ref{234para}. The best fit
of a polynomial logarithmic arm model for a spiral structure is presented in
Fig.~\ref{5arm8.0}, and the parameters are listed in Table~\ref{5arm}. The
number of arms and arm structure are conserved.  As occured for the case of
8.5/220, the tangential directions are better reproduced by the polynomial
model than the logarithmic spiral arm models.

\citet{rmz+09} used the newly measured trigonometric parallaxes of masers to
estimate the solar parameters, and concluded that $R_0=8.4\pm 0.6$~kpc and
$\Theta_0=254\pm16$ km s$^{-1}$. They offer a revised prescription (and a
Fortran code) to calculate kinematic distances and their uncertainties
\citep[see][for details]{rmz+09}. Using the code they kindly provided to
us, we determined the kinematic distances and re-scaled other parameters of
our HII regions and GMC samples. Their distribution and the best-fitted
polynomial logarithmic-arm model is shown in Fig.~\ref {5arm254}. Compared
with the distribution with $R_0=8.5$ kpc and $\Theta_0=220$ km s$^{-1}$, the
tracers are nearer to our sun in the second and third Galactic quadrants, so
that arms in this region are wound more tightly. The local arm becomes
clearer, but the separation between the Perseus arm and Perseus+1 becomes
less clear. The grand design of spiral arms in the other parts of the
Galaxy looks very similar to that of 8.0/220.

\section{Conclusions and remarks}

We collected the spiral tracer data of HII regions and giant molecular
clouds (GMCs) published in the literature, and used them to study the
weighted distribution of these tracers for the overall design of spiral arms
of our Milky Way Galaxy.  The mass of the GMC and the excitation parameter
of an HII region are scaled as the weighting factors. The distances of
tracers, if not available from stellar or triangulation measurements, were
estimated kinetically using the rotation curve of BB93 with solar constants
of $R_0=8.5$ kpc and $\Theta_0=220$ km s$^{-1}$. We fitted the distribution
of these tracers with two, three, and four logarithmic spiral arm
models. None of the models can match all the main arm features observed. We
finally fit the tracer data by using a polynomial logarithmic arm model,
which not only can fit the distribution of tracers for the main spiral arms,
but also match the tangential directions with the observations. The
polynomial model appears to be better able to describe the spiral arms of
our Galaxy. When the kinematic distances are estimated by using rotation
curves with two sets of parameters, $R_0=8.0$ kpc and $\Theta_0=220$ km
s$^{-1}$ and $R_0=8.4$ kpc and $\Theta_0=254$ km s$^{-1}$, the distribution
of HII regions and GMCs are slightly better fitted by the models, though the
parameters do not change significantly.

For most of tracers in our work, we use their kinematic distances derived
from the rotation curves. \citet{gom06} found that the error bars are
significantly large for objects at the positions of spiral arms. This
pitfall cannot be overcome until accurate distances of a large sample of HII
regions are measured, e.g. by triangulation of radio masers
\citep[e.g.][]{xrzm06,rmz+09}. A large project is being carried out with the
VLBA to directly measure distances of a large sample of massive star
formation regions by trigonometric parallaxes \citep[see][]{rmb+08} with
accuracies approaching $\sim 10 \mu$as.  Published results of several
examples show systematically smaller distances than the kinematic
distances. They will probably obtain hundreds of measurements over 10 years
for the brightest HII regions covering all the Galactic disk, and the final
grand design of our Milky Way will be greatly improved. However, at present,
the picture shown in Fig.~\ref{5arm254} is the best approximate description
of Galactic spiral structure.

\begin{acknowledgements}
We thank the referee for instructive comments, and Dr. Mark Reid
for providing the Fortran code for the estimation of kinematic
distances using their newly determined rotation curves. The authors
are supported by the National Natural Science Foundation (NNSF) of
China (10773016, 10821061 and 10833003) and the National Key Basic
Research Science Foundation of China (2007CB815403).
\end{acknowledgements}

\bibliographystyle{aa} % style aa.bst
\bibliography{Galac_structure.bib}

\appendix
\section{Tracer data for spiral structure: (online version only)}

We collected and calculated the related parameters of HII regions and
giant molecular clouds (GMCs) based on the observations in the
literature, and present them here in two tables.

\scriptsize
\setlength{\tabcolsep}{1.4mm}
\begin{longtable}{rrcrrrrrrrrrrrrrrrrrr}
\caption{\label{tab_a1} HII regions as tracers of spiral arms.
Columns 1 to 3 list the Galactic longitude, Galactic latitude and
velocity, taken from the reference given in Column 4; Column 5
to 7 give the radio continuum flux of the HII region,
observation frequency and the reference.  Column 8 to 10 give 
the stellar distance when available, as well as the error and the
reference; Column 11 is a note for the distance ambiguity
and Column 11 is its reference: Knear--the nearer kinematic distance is
adopted; Kfar--the farther kinematic distance is adopted; Ktan--the
tracer is located at the tangential point and the distance to the
tangent is adopted; Kout--the object is in the outer Galaxy and
only one kinematic distance is available and adopted; Column 13 and 14
give the kinematic distance to the Sun and distance to the Galactic
Center estimated with a rotation curve with $R_0=8.5$ kpc,
$\Theta_0=220$ km s$^{-1}$; Column 15 contains the excitation
parameter rescaled by the distance we adopted; Column 16 to 18 are the
same as Column 13-15 but with constants $R_0=8.0$ kpc and
$\Theta_0=220$ km s$^{-1}$ for kinematic distances. Column 19 to 21
are the same as Column 13-15 but with constants $R_0=8.4$ kpc and
$\Theta_0=254$ km s$^{-1}$ for kinematic distances.
References: ab09: \citet{ab09}; ahck02: \citet{ahck02}; bdc01:
\citet{bdc01}; brm+08: \citet{brm+08}; ccg+98: \citet{ccg+98}; cf08:
\citet{cf08}; cor+07: \citet{cor+07}; ch87: \citet{ch87}; gg76:
\citet{gg76}; GB6: \citet{gsdc96}; hh09: This work; kjb+03: \citet{kjb+03};
kb94: \citet{kb94}; lt08: \citet{lt08}; mrm+08: \citet{mrm+08}; pbd+03:
\citet{pbd+03}; pdd04: \citet{pdd04}; pmg08: \citet{pmg08}; rus03:
\citet{rus03}; rag07: \citet{rag07}; rmb+08: \citet{rmb+08}; swa+04:
\citet{swa+04}; tlhj08: \citet{tlhj08}; was+03: \citet{was+03}; PMN:
\citet{wgbe94,wgh+96,gwbe94,gwbe95}; xrm+08: \citet{xrm+08}; zzr+08:
\citet{zzr+08}.
} \\
\hline \hline 
l & b & $V_{lsr}$ & $Ref.$ & S & Freq & $Ref.$&
d & $d_{erro}$ & $Ref.$ & Mark & $Ref.$& $D_{8.5}$ & $R_{8.5}$ & $U_{8.5}$ &
$D_{8.0}$ & $R_{8.0}$ & $U_{8.0}$ & $D_{8.4}$ & $R_{8.4}$ & $U_{8.4}$
\\ 
$(^o)$& $(^o)$ & km/s & & Jy & GHz & &
kpc &kpc & & & &kpc& kpc & pc $cm^{-2}$ & kpc & kpc & pc $cm^{-2}$ & 
kpc & kpc & pc $cm^{-2}$ 
\\ (1) &
(2) &(3) & (4) & (5) & (6) & (7) & (8) & (9) & (10)& (11) & (12) & (13) &
(14) & (15) & (16) & (17) & (18) & (19) & (20) & (21) \\ 
\hline 
\endfirsthead
\caption{continued.}\\
\hline\hline
  l & b & $V_{lsr}$ & $Ref.$ & S  & Freq  & $Ref.$& d & $d_{erro}$  & $Ref.$
  & Mark & $Ref.$& $D_{8.5}$ & $R_{8.5}$  & $U_{8.5}$ & $D_{8.0}$ &
  $R_{8.0}$ & $U_{8.0}$ & $D_{8.4}$ & $R_{8.4}$ & $U_{8.4}$\\ 
  (1) & (2) &(3) & (4) & (5)  & (6)  & (7) & (8) & (9) & (10)& (11)  & (12) & (13) & (14)  & (15) & (16) & (17) & (18) & (19) & (20) & (21)\\
\hline
\endhead
\hline
\endfoot
  2.40 & 1.40 &   3.0 &pbd+03 &  3.00 & 2.70 &pbd+03 & 4.7   &1.4  & rus03 &       &       &      &      &  54.7 &      &      &  54.7 &      &     &   54.7\\
  2.90 &-0.00 &  -1.7 &pbd+03 &  3.40 & 2.70 &pbd+03 &       &     &       & kout  &  hh09 & 18.7 & 10.3 & 143.3 & 17.4 &  9.4 & 136.6 & 18.2 & 9.7 &  140.7\\
  4.30 & 0.60 &   6.6 & rus03 &       &      &       & 1.3   &0.4  & rus03 &       &       &      &      &  28.0 &      &      &  28.0 &      &     &   28.0\\
  6.00 &-1.30 &   2.8 &pbd+03 & 44.90 & 2.70 &pbd+03 & 1.6   &0.25 & rus03 &       &       &      &      &  65.8 &      &      &  65.8 &      &     &   65.8\\
  6.00 &-1.20 &   3.6 &pbd+03 &113.40 & 2.70 &pbd+03 & 1.6   &0.25 & rus03 &       &       &      &      &  89.5 &      &      &  89.5 &      &     &   89.5\\
  6.20 &-1.20 &   3.0 &pbd+03 &  2.10 & 2.70 &pbd+03 & 1.6   &0.25 & rus03 &       &       &      &      &  23.7 &      &      &  23.7 &      &     &   23.7\\
  6.40 &-0.50 &  19.0 &pbd+03 &  4.00 & 2.70 &pbd+03 &       &     &       & knear & pdd04 &  3.8 &  4.8 &  52.3 &  3.5 &  4.5 &  49.5 &  3.6 & 5.0 &   50.4\\
  6.50 & 0.10 &  14.1 &pbd+03 &  0.50 & 2.70 &pbd+03 &       &     &       & knear & pdd04 &  3.1 &  5.4 &  22.8 &  2.9 &  5.1 &  21.8 &  2.9 & 5.6 &   21.8\\
\hline\hline
\end{longtable}

\setlength{\tabcolsep}{2.0mm}
\begin{longtable}{rrrrrrrrrrrrrrrrr}
\caption{\label{tab_a2} GMCs as tracers of spiral arms. 
Columns 1 and 2 are the Galactic longitude and latitude; Column 3 to 6
list velocity, the luminosity of CO emission line (in unit of
$10^3$~K~km~s$^{-1}$~pc$^2$), distance and mass of molecular cloud,
which were all taken from the reference given in Column 8; Column 7 is
the note for distance ambiguity given in this reference, used as the 
same convention as in Table~A1; Columns 9 and 10
list the distances to the Sun and to the Galactic center, which are
estimated by the velocity and the rotation curve with $R_0=8.5$ kpc
and $\Theta_0=220$ km s$^{-1}$ when the stellar distance is
unavailable; Column 11 gives the mass of molecular clouds re-scaled by
the newly estimated distance; Columns 12, 13 and 14 are the same as
Columns 9-11 but with $R_0=8.0$ kpc and $\Theta_0=220$ km
s$^{-1}$. Columns 15, 16 and 17 are the same as Columns 9-11 but with
$R_0=8.4$ kpc and $\Theta_0=254$ km s$^{-1}$.
References: bw94: \citet{bw94}; css90: \citet{css90}; dect86:
\citet{dect86}; dbt90: \citet{dbt90}; dgt94: \citet{dgt94}; dlp+96:
\citet{dlp+96}; dkm+08: \citet{dkm+08}; gcbt88: \citet{gcbt88}; hcs01:
\citet{hcs01}; mk88: \citet{mk88}; mab97: \citet{mab97}; nomf05:
\citet{nomf05}; sod91: \citet{sod91}; srby87: \citet{srby87}; tllw07: \citet{tllw07}.
} \\ 
\hline
\hline 
l & b & $V_{lsr}$& $L_{CO}$ &D & $M_{GMCs}$ &Mark& Ref.& $D_{8.5}$ &
$R_{8.5}$ & $M_{8.5}$ & $D_{8.0}$& $R_{8.0}$& $M_{8.0}$ & $D_{8.4}$& $R_{8.4}$& $M_{8.4}$ \\
$(^o)$& $(^o)$& km/s& $10^3Kkms^{-1}pc^2$ & kpc& $10^5M_{\odot}$& & & kpc& kpc&
$10^5M_{\odot}$& kpc& kpc &$10^5M_{\odot}$ & kpc& kpc &$10^5M_{\odot}$ \\
(1) & (2) & (3)& (4) &(5) & (6)
&(7) & (8)& (9) & (10) & (11) & (12)& (13)& (14) & (15)& (16)& (17) \\ \hline \endfirsthead
\caption{continued.}\\
\hline\hline
  l & b & $V_{lsr}$& $L_{CO}$ &D & $M_{GMCs}$ &Mark& Ref.& 
$D_{8.5}$ & $R_{8.5}$ & $M_{8.5}$ & $D_{8.0}$& $R_{8.0}$& $M_{8.0}$ & $D_{8.4}$& $R_{8.4}$& $M_{8.4}$   \\
 (1) & (2) & (3)& (4) &(5) & (6) &(7) & (8)& 
(9) & (10) & (11) & (12)& (13)& (14)& (15)& (16)& (17)    \\ 
\hline
\endhead
\hline
\endfoot
  8.00 & -0.50 & 128.00 &   72.70 &10.10 &   4.435 &ktan  &srby87 &     &      &        &     &      &        &      &     &      \\
  8.20 &  0.20 &  20.00 & 1402.00 &15.90 &  85.522 &kfar  &srby87 &13.4 &  5.2 &  60.74 &12.7 &  4.9 &  54.56 & 13.4 & 5.2 & 60.74\\
  8.30 & -0.10 &  48.00 &   50.20 &13.20 &   3.062 &kfar  &srby87 &11.5 &  3.3 &   2.32 &10.9 &  3.2 &   2.09 & 11.6 & 3.4 &  2.36\\
  8.30 & -0.30 &  16.00 &   28.10 & 3.20 &   1.714 &knear &srby87 & 2.9 &  5.6 &   1.41 & 2.7 &  5.3 &   1.22 &  2.8 & 5.8 &  1.31\\
  8.40 & -0.30 &  37.00 &  233.00 & 5.70 &  14.213 &knear &srby87 & 4.7 &  3.9 &   9.66 & 4.4 &  3.7 &   8.47 &  4.4 & 4.1 &  8.47\\
  8.50 & -1.00 &  16.00 &  106.00 & 3.20 &   6.466 &knear &srby87 & 2.9 &  5.7 &   5.31 & 2.7 &  5.4 &   4.60 &  2.7 & 5.8 &  4.60\\
  8.70 &  0.60 &  22.00 &    8.60 & 4.00 &   0.525 &knear &srby87 & 3.5 &  5.1 &   0.40 & 3.2 &  4.8 &   0.34 &  3.3 & 5.3 &  0.36\\
  9.30 &  0.70 &  15.00 &    2.00 & 2.80 &   0.122 &knear &srby87 & 2.6 &  6.0 &   0.11 & 2.4 &  5.6 &   0.09 &  2.4 & 6.1 &  0.09\\
\end{longtable}

\end{document}